\documentstyle[referee]{aa}
\input psfig.tex
\def\lsim{\lower.5ex\hbox{$\; \buildrel < \over \sim \;$}}
\def\gsim{\lower.5ex\hbox{$\; \buildrel > \over \sim \;$}}

.tfm

\begin{document}

\title{Signatures of accretion shocks in emitted radiation
from a two temperature advective flow around black holes}
\author {Samir Mandal$^{1}$ \and Sandip K. Chakrabarti$^{2,1}$ }
\institute{$^1$ Centre for Space Physics, Chalantika 43, Garia Station Rd., \\
Garia, Kolkata, 700084, e-mail: space\_phys@vsnl.com\\
$^2$ S.N. Bose National Centre for Basic Sciences,\\
JD Block, Salt Lake, Sector III, Kolkata 700098, e-mail: chakraba@bose.res.in}
\offprints{S. Mandal,  {\it space\_phys@vsnl.com}}
\date{Received ; accepted , }
\markboth{  }{}

\abstract{
Centrifugal barrier supported boundary layer (CENBOL) of a black 
hole affects the spectrum exactly in the same way the boundary 
layer of a neutron star does. The CENBOL is produced due to standing 
or oscillating shock waves and these shocks accelerate electrons 
very efficiently and produce a power-law distribution. The accelerated 
particles in turn emit synchrotron radiation in presence of the 
magnetic field. We study the spectral properties of an accretion 
disk as a function of the shock strength, compression ratio, 
flow accretion rate and flow geometry. In the absence of a satisfactory 
description of magnetic fields inside the advective disk, we 
consider the presence of only stochastic fields and use the 
ratio of the field energy density to the gravitational energy density to be
a parameter. Not surprisingly, stronger fields produce stronger humps
due to synchrotron radiation. We not only include `conventional'
synchrotron emission and Comptonization due to Maxwell-Bolzmann 
electrons in the gas, we also compute these effects due to power-law 
electrons. For strong shocks, a bump is produced at a frequency 
just above the synchrotron self-absorption frequency 
at $\nu_{bump} \sim \nu_{inj} [1+\frac{4}{3}\frac{R-1}
{R}\frac{1}{x_s^{1/2}}]^{x_s^{1/2}}$, where, $\nu_{inj}$ is the 
frequency of the dominant photons from the pre-shock flow, $R$ 
is the compression ratio of the shock located at $x_s$. For strong 
shocks, a bump at a higher frequency appears predominantly due to
the power-law electrons formed at the shock front. }

\titlerunning{Signatures of accretion shocks in emitted radiation}
\authorrunning{Mandal and Chakrabarti}
\maketitle

\noindent ASTRONOMY AND ASTROPHYSICS (IN PRESS)

\section{Introduction}

Attempts to explain the spectra of black holes were made in the 
early seventies (Shapiro, 1973ab) by using spherically symmetric Bondi 
flows (Bondi, 1952).  However, the flow was found to be very radiatively 
inefficient due to high advection and in the same year Shakura-Sunyaev 
(1973) proposed a disk model in which the angular momentum distribution 
is kept as the Keplerian distribution around a black hole. However, 
the emitted multi-colour black body radiation was not capable of explaining 
the spectra of Cyg X-1 and Eardley, Lightman, \& Shapiro, (1975); Shapiro, 
Lightman \& Eardley (1976) showed that it is possible to have a `hotter' 
branch of the solution, still retaining Keplerian distribution, which can 
produce hard X-rays from the inner region. On the other hand, Eardley \& 
Lightman (1975) found that the Keplerian disk with a constant $\alpha$ was 
unstable due to thermal and viscous effects (1975) which was 
later  corroborated by Eggum, Coroniti \& Katz (1985) by detailed numerical 
simulation.  They showed that the Keplerian disk collapsed in presence of 
constant viscosity parameter.  The focus was then to introduce a mixture of
sub-Keplerian and super-Keplerian (non-accreting) flows, such as thick 
accretion disk (Paczy\'nski \& Wiita, 1980) in the optically thick limit,
or accretion torus (Rees et al. 1982) in the optically thin limit and their 
spectral properties were studied (e.g., Madau, 1988). Other models included
`non-Keplerian' components such as, magnetic corona (Galeev, Rosner \& Vaiana  
1979). At the same time, the sub-Keplerian transonic flow models were introduced 
(Muchotrzeb \& Paczy\'nski, 1982) and they were improved upon to include standing,
oscillating or propagating shock waves (e.g. Chakrabarti, 1989; Chakrabarti \& 
Molteni, 1995; Molteni, Sponholz \& Chakrabarti, 1996). Subsequently,
Chakrabarti \& Titarchuk (1995, hereafter CT95) incorporated this flow solution in computing 
spectra of black holes. In recent years, it has become customary to include the
sub-Keplerian flows in accretion flow models because of the
compelling evidences that the Keplerian disk alone cannot 
explain most of the observational facts  and two components were essential (Smith, Heindl \& 
Swank, 2002; Smith, Heindl, Markwardt and Swank, 2001;
see Chakrabarti, 1996 for a review). There are several other models
in the literature which also include a secondary component 
such as Compton cloud or magnetic corona (Poutanen \& Coppi, 1998; Gierli\'nski et al. 1999;
Coppi, 1999; Zdziarski et al. 2001) which also satisfactorily fit spectra till high energies.
Esin et al. (1998) fitted spectra with ADAF model in which the Keplerian 
flow becomes ion pressure dominated in the inner region. In the present paper,
we shall concentrate on spectral properties of the dynamic `Compton cloud', namely,
the sub-Keplerian transonic flow component and not deal with other models present in the 
literature. 

In CT95, a two component advective flow (TCAF) solution was introduced 
in which the Keplerian disk was on the equatorial plane as usual, but a dynamic, 
sub-Keplerian halo component (basically the transonic solutions worked out
earlier, see, Chakrabarti 1996) flanks this Keplerian disk. While the Keplerian 
disk was supplying soft-photons, the sub-Keplerian flow was supplying hot electrons 
and the relative importance of these two accretion rates determined whether the 
spectrum was going to be soft or hard. A pivotal role is played in this context 
by the CENtrifugal barrier supported BOundary Layer or CENBOL. This is the hot and 
the puffed-up, post-shock region which intercepts soft-photons from the Keplerian 
disk and reprocesses them through inverse Comptonization. It is to be remembered 
that the CENBOL is a direct consequence of the dominance of the centrifugal force 
near a black hole and it is not to be confused with puffed up corona or magnetized 
corona that is used by other models. We find it comfortable to use CENBOL as 
the Comptonizing cloud as its existence and stability have been shown beyond doubt in 
a large number of numerical simulations (Molteni, Lanzafame \& Chakrabarti, 1994;
Chakrabarti \& Molteni, 1995; Molteni, Sponholz, \& Chakrabarti, 1996).
(Non-axisymmetric perturbations could not remove them also, 
see, Molteni, T\'oth \& Kuznetsov, 1999). We can produce the 
observed power-law component without taking resort to any 
hypothetical electron cloud. In essence, the CENBOL region in between 
the horizon and the shock behaves like a boundary layer where 
the flow dissipates its gravitational energy. This boundary layer 
can oscillate when the cooling is introduced  and this is proposed 
(Molteni, Sponholz \& Chakrabarti, 1996; Chakrabarti \& Manickam, 2000; 
Chakrabarti, Acharyya \& Molteni, 2004) as the cause of the Quasi-Periodic Oscillation.

Given that a significant fraction of the accretion flow could pass through a 
standing, oscillating or a propagating shock, it is likely that the hot 
electrons may be {\it accelerated} by them just as the high energy cosmic 
rays (see, Kazanas \& Elision, 1986, Molteni \& Sponholz, 1994, Chakrabarti, 1996) 
are produced by the transient super-novae shocks 
(Bell 1978ab, Longair 1981). In the present analysis, we shall incorporate 
the power-law component of the electrons to compute the synchrotron radiation 
and the Comptonization of these photons by these electrons. The inclusion of 
the synchrotron emission generated by non-thermal  electrons is not new. For example,
Lin \& Liang (1999) considered a hybrid model which included a few percent of 
non-thermal electrons along with the usual thermal electron obeying Maxwell-Boltzmann
distribution. The emission included thermal and magnetic bremsstrahlung but no
Comptonization. In a parametric study, with the percentage of 
non-thermal electrons as a free parameter, Wardzinski \& Zdziarski (2001)
showed that the soft photons generated could significantly change the spectra. These
works do not address the origin of the non-thermal electrons quantitatively but propose
that they could be produced in magnetized corona. In contrast, we assume that {\it all} the 
non-thermal electrons are produced in the accretion shocks, though, 
some electrons from magnetized corona could not be ruled out and these should be included
for a completely consistent solution. Some of the examples of changes of spectral states of 
the black holes are presented in Mandal \& Chakrabarti (2004).

In the next Section, we present the basic equations and discuss the relevant 
parameters for the problem. We assume the flow to be around a
Schwarzschild black hole.  For the sake of simplicity, we use
the well-known Paczy\'nski \& Wiita (1980) pseudo-Newtonian potential 
($\phi_{PN}=-\frac{GM}{r-r_g}$, where $r_g=2GM/c^2$ is the Schwarzschild radius of the
black hole, $M$ is the mass of the black hole, $G$ and $c$ are the 
universal gravitational constant and the velocity of light respectively.)
to describe the black hole geometry. 
In Section 3, we discuss the nature of heating and cooling processes in an advective flow.
In Section 4, we present our procedure to compute the Comptonized 
spectra using power-law electrons. In Section 5, we explain the solution
procedure. In Section 6, we present the emergent spectra for various parameters and 
discuss their significance. Finally, in Section 7, we present concluding remarks.

\section{Basic hydrodynamic equations}

We consider a wedge shaped, axi-symmetric, low angular momentum flow (Chakrabarti 1990)
having the number density distribution:
$$
n(x) = \frac{\dot M}{\Omega m_p x^2 v(x)}.
\eqno{(1)}
$$
Here, $ \Omega $ is the solid angle subtended by the flow:
$$
\Omega=4\pi cos (\Theta) ,
\eqno{(2)}
$$
$\Theta$ is the angle made by the surface of the flow with the vertical axis, $m_p$
is the mass of the proton, $x$ is the radial distance $r$ in units of $r_g$ and ${\dot M}$
is the accretion rate. The radial velocity, 
$$
v(x)\sim (x-1)^{-1/2}
\eqno{(3)}
$$ 
increases rapidly in this sub-Keplerian flow. In presence of pure hydrogen, 
Thomson scattering will be the most dominating scattering process and the 
corresponding optical depth $ \tau(x)$ can be computed using  Eqs. (1) and (3).
In the absence of any satisfactory description of the magnetic field in accretion disks,
we assume only a random or stochastic field of strength $B(r)$. 
This may or may not be in equipartition with the
ionized flow. We define a parameter $\beta$ which is the ratio of the magnetic energy
density and the gravitational energy density:
$$
\beta=\frac{B^2 (r-2GM/c^2) }{8\pi GM\rho}.
\eqno{(4)}
$$
Normally, $\beta\lsim 1$ and for $\beta>1$, the magnetic field may buoyantly 
leave the disk.

The ions are expected to supply Coulomb energy to the electrons which in turn
become cooler using thermal and magnetic bremsstrahlung processes. In the presence of 
soft photons, they can loose energy via inverse Compton processes. 

The energy balance equation for the protons and electrons in our model could be
written as:

$$
\frac{dT_p}{dx} + \frac{T_p(3x-4)}{3x(x-1)} + \frac{\Omega m_p}{k\dot M} 
\frac{2}{3}x^2(\Gamma_p - \Lambda_p) = 0,
\eqno{(5a)}
$$
$$
\frac{dT_e}{dx} + \frac{3}{2}(\gamma_{pol} - 1)\frac{T_e(3x-4)}{3x(x-1)} + 
\frac{\Omega m_p}{k\dot M}(\gamma_{pol} - 1)x^2(\Gamma_e - \Lambda_e) = 0,
\eqno{(5b)}
$$
where, $\gamma_{pol} $ is $5/3$ for non-relativistic electron temperatures 
$(T_e \leq m_e c^2/k)$ and $4/3$ for relativistic electron temperatures
$(T_e > m_e c^2/k)$. $k$ is the Boltzmann constant. Since protons are
much heavier than the electrons, $T_p$ always remains in the non-relativistic domain. 
$ \Gamma $ and $ \Lambda $ contain contributions from all the heating
and cooling processes respectively. We shall describe various terms shortly below.

One can integrate these equations simultaneously using a fourth order Runge-Kutta method.
We impose a condition that a steady accretion shock occurs at $x=x_s$ through which 
the energy flux and the mass flux are conserved.  
In Chakrabarti (1989) it was shown that the shocks in a Schwarzschild geometry 
typically occur at around $x_s\sim 10-80r_g$ depending on specific angular 
momentum $\lambda$. Instead of using $\lambda$ as a free parameter
we use $x_s$ to be the free parameter. We use $R$, the compression ratio, 
to be a free parameter as well. Normally, in a transonic solution the shock 
location and the compression ratio are obtained simultaneously (Chakrabarti 1989).

\section{Nature of the heating and cooling Processes} 

Protons lose energy through Coulomb interaction $\Lambda_{ep}$ and 
inverse bremsstrahlung $\Lambda_{ib}$:
$$
\Lambda_p = \Lambda_{ep} + \Lambda_{ib}.
\eqno{(6)}
$$
Here, the subscript $p$ stands for protons. Electron-proton coupling supplies 
energy to the electrons (from protons) through $\Lambda_{ep}$ which 
it lost due to various cooling processes. This behaves as a heating term as far as the 
electrons are concerned and is given by:
$$
\Lambda_{ep} = 1.6 \times 10^{-13} \frac{k {\sqrt m_e} ln \Lambda_0}{m_p} n^2 
(T_p - T_e) T_e^{-3/2},
\eqno{(7)}
$$
where, $ln \Lambda_0$ is the Coulomb logarithm, $m_p$ and $m_e$ are the rest masses of
protons and electrons respectively. Electrons are heated through this Coulomb coupling.
$$
\Gamma_e = \Lambda_{ep}.
\eqno{(8)}
$$
Subscript $e$ stands for electrons. 

Electrons are cooled by bremsstrahlung at the rate of $\Lambda_{b}$, 
by cyclo-synchrotron  at the rate of $\Lambda_{cs}$ and by Comptonization at the
rate of $\Lambda_{mc}$ of the soft photons due to cyclo-synchrotron radiation. 
The net cooling of the electrons is:
$$
\Lambda_{e} = \Lambda_{b} + \Lambda_{cs} + \Lambda_{mc}  .
\eqno{(9)}
$$
Explicit expressions for these cooling terms for electrons satisfying
Maxwell-Bolzmann (MB) distribution are:
$$
\Lambda_{ib} = 1.4 \times 10^{-27} n^2 \Bigl(\frac{m_e}{m_p} T_p \Bigr)^{1/2} ,
\eqno{(10a)}
$$
$$
\Lambda_{b} = 1.4 \times 10^{-27} n^2 T_e^{1/2}(1+4.4 \times 10^{-10} T_e) ,
\eqno{(10b)}
$$
$$
\Lambda_{cs} = \frac{2\pi}{3c^2} kT_e(x) \frac{\nu_a^3}{x}, 
\eqno{(10c)}
$$
where $\nu_a$ is the critical frequency at which the self-absorbed synchrotron
radiation spectrum is peaked and it can be determined from the relation,
$$
\nu_a = \frac{3}{2} \nu_0 \theta_e^2 x_m ,
\eqno{(11)}
$$ 
where,
$$
\nu_0 = 2.8 \times 10^6 B,
\eqno{(12a)}
$$
$$
\theta_e = \frac{k T_e}{m_e c^2}.
\eqno{(12b)}
$$
Procedure to determine $x_m$ will be discussed below. 

So far, we discussed the situation when the electrons obey MB
velocity distribution law.  However, in our present situation, a section of the flow passes through the 
shock and a fraction $\zeta$ of the electrons will have a power-law distribution. For instance, the cooling
term due to cyclo-synchrotron photons would be given by (Longair 1981),
$$
\Lambda_{cs}={\cal A} {\cal G} B^{(p+1)/2} (\nu_{max}^{(3-p)/2} - \nu_{min}^{(3-p)/2})
\eqno{(13)}
$$
where, 
$$
{\cal A}= \frac{(3\pi)^{1/2}K e^3 } {m_e c^2 (1+p)(3-p)} (\frac{2\pi m_e^3 c^5}{3e})^{(1-p)/2} ,
\eqno{(14a)}
$$
$$
{\cal G}= \frac{\Gamma(p/4+19/12)\Gamma(p/4-1/12)\Gamma(p/4+5/4)}{\Gamma(p/4+7/4)} .
\eqno{(14b)}
$$
and $K$ is the normalization constant of power-law electron distribution,
$$
n({\cal E})|_{x_s}=K{\cal E}^{-p},
\eqno{(15)}
$$ 
which is obtained using the constraint that the electron number is conserved 
during the shock acceleration. Here $\nu_{min}$ is the minimum frequency
considered for integration. When self-absorption is included, this may 
be treated as $\nu_{a}$ defined above. The Comptonization is computed 
by using this cooling term augmented by the enhancement factor 
${\cal F}$:
$$
\Lambda_{mc} = \Lambda_{cs} {\cal F},
\eqno{(16)}
$$
For MB distribution, ${\cal F}$ is given by,
$$
{\cal F}= \eta_1 \Bigl \{1 - \Bigl(\frac{x_a}{3\theta_e} \Bigr)^{\eta_2} \Bigr \},
\eqno{(17)}
$$
where, 
$$
\eta_1=\frac{P(A-1)}{(1-PA)},
\eqno{(18a)}
$$ 
$$
P=1-exp(-\tau_{es}),
\eqno{(18b)}
$$ 
is the probability that an escaping photon is scattered.
$$
A=1+4\theta_e+16\theta_e^2,
\eqno{(19)}
$$ 
is the mean amplification factor in the energy of a scattered photon when the 
scattering electrons have a Maxwellian velocity distribution of temperature 
$\theta_e$, $\eta_2=1-\frac{lnP}{lnA}$ and $x_a=h\nu_a/m_e c^2$.
The seed photons generated by the electrons obeying power-law distribution
will also be comptonized by the thermal electrons and
the amplification factor (Dermer, Liang \& Canfield, 1991) can be written as,
$$
{\cal F}=\eta_1 [1-(\frac{s}{\phi_s}) \frac{x_{max}^{\phi_s}-x_a^{\phi_s}}
{3 \theta_e(\phi-1)(x_{max}^s - x_a^s}],
\eqno{(20)}
$$
where, 
$$
x_{max}=\frac{h\nu_{max}}{m_e c^2},
$$ 
$$
s=\frac{5-p}{2},
$$
$$ 
\phi_s=\phi +\frac{3-p}{2},
$$ 
and 
$$
\phi=\eta_2-1.
$$  
Similarly, the amplification factor (Rybicki \& Lightman 1979) for the seed photons Comptonized
by the electrons obeying power-law distribution can be written as,
$$
{\cal F}=\frac{4}{3}{{\sigma}_{T}} R_{c} K \frac{({\cal E}_{max}^{3-p}-{\cal E}_{min}^{3-p})}{(3-p)},
\eqno{(21)}
$$
where $R_{c}$ is the size of the Comptonized region. ${\cal E}_{max}$ and ${\cal E}_{min}$ 
are maximum and minimum energy of the power-law electrons respectively.

In the current situation both the thermal and the non-thermal electrons are present. 
In this case, $x_m$ is calculated by equating the total loss due to the thermal 
and the non-thermal emission from the CENBOL with the contributions
from the corresponding source functions from the CENBOL surface at the
self-absorption frequency $\nu=\nu_a$, i. e.,
$$
[\epsilon(\nu)_{MB} + \epsilon(\nu)_{PL}].V = \pi[(1-\zeta).p(\nu)_{MB}+
\zeta.p(\nu)_{PL}].S,
\eqno{(22)}
$$
where, 
$$
\epsilon(\nu)_{MB}=4.43 \times 10^{-30} \frac{4 \pi n {\nu}}{K_2(1/{\theta_e})}
I(\frac{x_m}{sin(\psi )}).
\eqno{(23)}
$$
$x_m$ is calculated from eq. (11):
$$
x_m=\frac{2 \nu}{3 \nu_0 {\theta_e}^2},
\eqno{(24a)}
$$
and $\psi$ is the angle between the velocity vector of the electron and the
direction of the local magnetic field. Averaging over $\psi$ for an isotropic
velocity distribution, $I(\frac{x_m}{sin(\psi )})$ gets replaced by 
$I^\prime ({x_m})$ and is given by (Narayan \& Yi, 1995),
$$
I^\prime ({x_m})=\frac{4.0505}{{x_m}^{1/6}} (1+\frac{0.40}{{x_m}^{1/4}}
+ \frac{0.5316}{{x_m}^{1/2}})  exp(-1.8899 {x_m}^{1/3}).
\eqno{(24b)}
$$
This is the emissivity of the electrons obeying MB distribution.
The emissivity of the electrons obeying power-law distribution is given by,
$$
\epsilon(\nu)_{PL}= \frac{3-p}{2} {\cal A} {\cal G} B^{(p+1)/2} (\nu)^{(1-p)/2}.
\eqno{(25)}
$$ 
The source function for thermal emission in the low frequency regime is given by,
$$
p(\nu)_{MB}=2 m_e {\nu}^2 \theta_e.
\eqno{(26)}
$$
The source function of power-law emission is given by,
$$
p(\nu)_{PL}= \frac{\sqrt {\pi}}{c^2(1+p)}(\frac{2\pi m_e^3 c^5}{3e})^{1/2}
{\cal H} B^{-1/2} \nu^{5/2},
\eqno{(27a)}
$$
where,
$$
{\cal H}= \frac{\Gamma(p/4+19/12)\Gamma(p/4-1/12)\Gamma(3/4)}{\Gamma(p/4+11/6)
\Gamma(p/4+1/6)\Gamma(5/4)}.
\eqno{(27b)}
$$
$\zeta$ is the percentage of electrons acquiring a power-law energy
distribution. In Eqn. (22), $V$ is the volume element under consideration 
and $S$ is the corresponding surface area.

We ignore the effects of power-law electrons on the emission of 
bremsstrahlung radiation since this radiation is very weak. 

\section{Computation of the spectral index}

In the previous Section, we mentioned how the injected photon intensity is
enhanced due to the inverse Comptonization process. In this Section,
we discuss how the spectral slope is computed.  In CT95, the electrons in 
the pre-shock and the post-shock regions were assumed to be following
Maxwell-Bolzmann distribution. Hence the spectral index was computed 
from Titarchuk \& Lyubarskij (1995).

The electrons from the pre-shock flow will be 
accelerated  at the shock and a fraction of post-shock electrons 
will obey the power-law distribution.
So, the CENBOL has both the MB as well as the power-law electrons. The energy spectral 
index ($\alpha$) of the outgoing radiations due to Comptonization 
depends on the distribution function of electrons. We follow the 
prescription suggested in Titarchuk \& Lyubarskij (1995) and Gieseler 
\& Kirk (1997) to calculate $\alpha$. A brief discussion about the
procedure of calculating $\alpha$ is given below. 

It is well known that a spectrum can be characterized by the dimensionless 
electron temperature $\theta_e=kT_e/m_e c^2$ and the optical depth $\tau$ 
of the medium (Sunyaev \& Titarchuk 1985). The condition that has to be satisfied is:
$$
C_0(\theta_e,\alpha)\lambda(\tau_0)=1.
\eqno{(28)}
$$
Here, $\lambda(\tau_0)=exp(-\xi)$, and 
$$
\xi=\frac{\pi^2}{3(\tau_0+2/3)^2}(1-exp(-.7\tau_0) + exp(-1.4\tau_0)\ln\frac{4}{3\tau_0},
\eqno{(29)}
$$
for spherical geometry and $\tau_0$ is the total optical depth of the CENBOL.
The quantity $C_0$ itself is to be obtained from,
$$
C_0(\theta_e,\alpha)=3\pi \int_0^1{v^2 dv\frac{f(v)}{\gamma^2}\hat C_0},
\eqno{(30)}
$$
where $v$, $f(v)$ and $\gamma$ are the velocity of electrons in units of the
velocity of light, electron velocity distribution function and the Lorentz factor 
respectively and 
$$
\hat C_0=\frac{3}{8\gamma^2}[I_{0,0}+I_{2,2}-\frac{1}{3}(I_{2,0}+I_{0,2})]. 
\eqno{(31)}
$$
The tabulated integrals $I_{i,j}$ is given by,
$$
I_{i,j}=\int_{-1}^1(1+vx)^{\alpha}x^i dx \int_{-1}^1
\frac{y^j dy}{(1+vy)^{\alpha+3}}.
\eqno{(32)}
$$
In the case of MB distribution,
$$
f(v)=\frac{\gamma^5 exp(-\gamma/\theta_e)}{4\pi \theta_e K_2(1/\theta_e)},
\eqno{(33a)}
$$
where $K_2(1/\theta_e)$ is the modified Bessel function of second kind
of order two and for power-law distribution (eq. 15),
$$
f(v)=\frac{1-p}{4\pi(\gamma_{max}^{1-p}-\gamma_{min}^{1-p})} v{\gamma^{3-p}} .
\eqno{(33b)}
$$
Here, $\gamma_{max}$ and $\gamma_{min}$ are obtained from particle energy
as will be discussed in next section.
Integrating over different $f(v)$ we get different $\alpha$ for the same set of flow
parameters. Using these $\alpha$ we calculate the spectrum from CENBOL in 
the same way as described in CT95.   

\section{Solution Procedure}

The geometry of the flow is chosen to be conical and the flow surface
makes an angle $\Theta$ with the z-axis. We use the Fourth-order 
Runge-Kutta method to integrate the energy equations 5(a-b). 
These two first order differential equations required two initial conditions 
when the flow geometry (determined mainly by the momentum equations) and the 
accretion rate (determined by the continuity equation) are provided. 
At the shock, $x=x_s$, the velocity $v(x)$ is reduced by a factor $R$, the compression 
ratio, i.e., $v_+(x_s)=v_-(x_s)/R$. The number density as obtained from Eq. (1) 
also goes up by a factor $R$. 

As for the initial conditions, we fix the outer boundary at a large distance 
(say, $10^6 r_g$) and supply matter (both electrons and protons) with the 
same temperature (say, $T_p=T_e=10^6K$). During integration, we obtained
temperature and radiation emitted by the flow through bremsstrahlung and 
synchrotron radiation. These low energy photons are then inverse Comptonized 
by the hot electrons in the flow.

A shock of compression ratio $R$ causes the formation of power-law 
electrons of slope (Bell, 1978ab),
$$
p=(R+2)/(R-1).
\eqno{(34)}
$$
This power-law electrons produce a power-law synchrotron
emission with an index $q$ given by (Longair, 1981):
$$
q=(1-p)/2.
\eqno{(35)}
$$
The power-law electrons have energy minimum at 
$$
{\cal E}_{min} =m_e c^2 \gamma_{min}
\eqno{(36a)}
$$
and energy maximum at 
$$
{\cal E}_{max}=m_e c^2 \gamma_{max} .
\eqno{(36b)}
$$ 
The ${\cal E}_{max}$ was obtained
self-consistently by conserving the number of power-law electrons
and by computing the number of scatterings that the electrons undergo
inside the disk before they escape. This yields,
$$
\gamma_{max}=\gamma_{min}[1+\frac{4}{3}\frac{R-1}{R}\frac{1}{x_s^{1/2}}]^{x_s^{1/2}}.
\eqno{(37)}
$$
The minimum value of the Lorentz factor $\gamma_{min}$ is obtained
from the temperature of the injected electrons. This temperature 
is obtained self-consistently through our integration procedure.
While obtaining the net cooling, we have used this local temperature and the local optical depth.
To compute the spectrum, we divided the pre-shock region into several annuli and the post-shock
region has been kept as a single zone. The number of intercepted soft-photons which
participated in the Comptonization process has been calculated using local optical depth.

To obtain a realistic idea of how much matter actually passes through the shock, we take resort to
the numerical simulations of Molteni, Lanzafame \& Chakrabarti (1994). This fraction
will depend on the angular momentum: for lower angular momentum, the flow passes close
to the equatorial plane and passes through the shock formed very close to the hole. Very 
little matter is lost as winds.  For higher angular momentum, the CENBOL size increases 
and a profuse amount of wind/outflow is formed after being processed through 
the CENBOL. The fraction can also be time-dependent. In the absence of a fully 
time dependent solution, we assume the percentage of electrons $\zeta$ acquiring 
a power-law energy distribution to be a free parameter. The goal is to obtain 
as varied spectra as possible in order to constrain the parameters of a system.

In presence of the  magnetic field, the pre-shock flow and the CENBOL would emit the 
synchrotron radiation. While computing this we have incorporated the synchrotron 
self-absorption by the emitting medium when it becomes optically thick.
Eqn. (11) gives the expression for the self-absorption frequency $\nu_a$.

Though there are several approaches in the literature 
(see, e.g., Poutanen \& Coppi, 1998; Gierli\'nski et al. 1999; 
Coppi, 1999; Zdziarski et al. 2001) we followed the procedures presented in 
CT95 and Titarchuk \& Lyubarskij (1995) while computing the Comptonized 
spectrum due to the scattering from MB electrons and the power-law electrons. However, unlike CT95 
where a Keplerian  disk was supplying soft photons, here the source is 
everywhere, i.e., distributed throughout. This has been taken into account.
At the end, we add contributions from all the components to get 
the net photon emissions from the flow. 

In the next Section, we present the spectral characteristics of the wedge shaped thin 
flow around a $10M_\odot$ black hole (unless otherwise stated)
having $\Theta=75^o$. We generally use the strong shock $R=4$ except in Fig. 4,
(where $R$ itself is varied) and saturated magnetic field $\beta=1$ 
except in Fig. 7 (where $\beta$ is varied). We vary other solution parameters: $x_s$, $\zeta$ 
and ${\dot m}$ to obtain results for different cases. Here, ${\dot m}$ is the mass 
accretion rate in units of Eddington rate ${\dot M}_{Edd}$. We generally use a 
black hole of mass $10M_\odot$. However, we probed the results when the mass 
was chosen to be $10^6M_\odot$ as well.

\section{Results and Interpretations}

First, we show the temperature distribution of the incoming matter in presence of
cooling. Figures 1(a-b) show the variation of the electron and proton
temperatures ($T_e$ and $T_p$ respectively) as functions of 
the radial distance $x$ (measured in units of the Schwarzschild 
radius $r_g$). Both the axis are in logarithmic scale. Solid curves show
the proton temperatures. 

In Fig. 1a, results for ${\dot m}=0.01$ $M=10 M_\odot$ (long-dashed curve of $T_e$), 
and $M=10^6M_\odot$ (dotted curve of $T_e$) are shown. The other parameters are: $\zeta=0.05$ 
and $x_s=35$. For the higher mass, the density is lower and electrons are less efficiently cooled.
In Fig. 1b, ${\dot m}=0.1$ (long-dashed curve of $T_e$) and ${\dot m}=0.5$ (dotted curve).
Other parameters are $M=10M_\odot$, $\zeta=0.2$, and $x_s=25$.
Higher accretion rate causes higher cooling of the electrons and thus $T_p$ and $T_e$ 
start to deviate from each other farther from the black hole. Second, the initial deviation
from proton temperature is also higher for higher rates. It is generally observed
that just after the shock, electrons 
of the flow with lower accretion rate cool faster. This is because the CENBOL has more synchrotron 
photons to cool fewer electrons. Thus the role of the CENBOL is crucial to change the nature 
of cooling quantitatively.

\begin {figure}
\vbox{
\vskip -3.0cm
\hskip 0.5cm
\centerline{
\psfig{figure=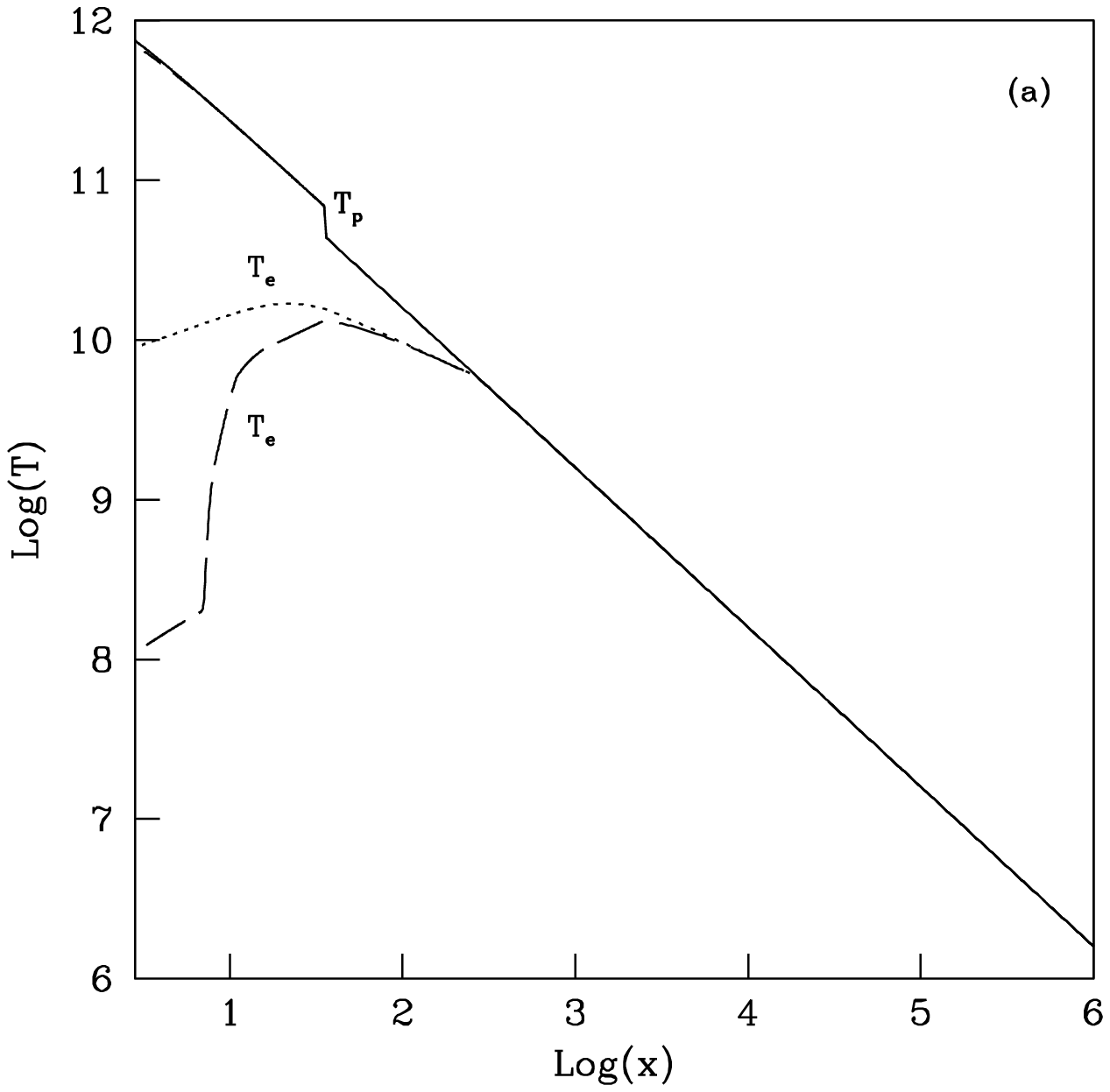,height=8truecm,width=8truecm}
\hskip -1.0cm
\psfig{figure=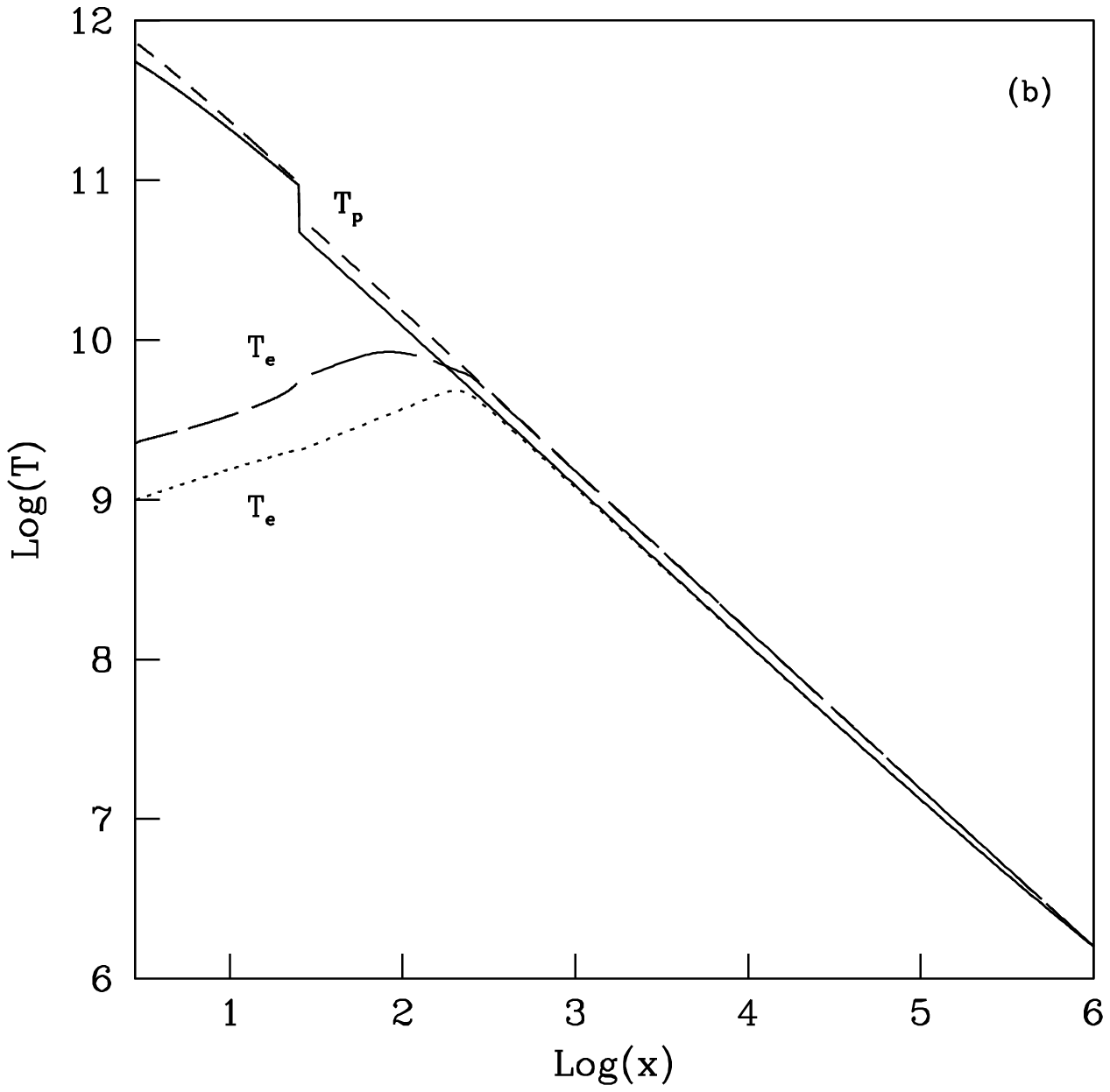,height=8truecm,width=8truecm}}}
\vspace{0.0cm}
\caption{
Variation of the proton (solid) and electron temperatures as  
functions of dimensionless radial distance $x$ in presence of a strong shock ($R=4$)
with saturated magnetic field ($\beta=1$). In (a), ${\dot m}=0.01$ 
and $M=10M_\odot$ (long-dashed) and $M=10^6M_\odot$ (dotted).
Other parameters are  $x_s = 35$, $\zeta=0.05$.
For higher mass, the density becomes lower and the electrons are less efficiently
cooled. In (b) ${\dot m}=0.1$ (long-dashed) and ${\dot m}=0.5$ (dotted). Other parameters are 
$M=10M_\odot$, $\zeta=0.2$ and $x_s=25$. For higher accretion rates, the pre-shock 
flow cools faster and the temperatures deviate from each other farther away.
}
\end{figure}

We now present a typical spectrum of an accretion flow around a $M=10M_\odot$ black hole (Fig. 2a) 
and around a $M=10^6M_\odot$  black hole (Fig. 2b). Other parameters are exactly the same as in Fig. 1a.
The curve marked 1 is due to the synchrotron emission  by the  thermal electrons from the pre-shock 
region and that marked 2 is due to the Comptonization of these photons.
These curves were drawn after the self-absorption was taken into account.
The curves marked 3 and 4 are due to the synchrotron emission by the thermal electrons and the
Comptonization of these photons by the same in the post-shock region respectively.
The curve marked 5 represents the synchrotron emission by the power-law electrons in the 
post-shock region. The curve marked 6 is due to Comptonization of the power-law generated 
soft photons. The curve marked 7 is the total spectrum emitted from the entire flow. 
The bumps in the spectrum at around $Log(\nu)\sim 14-16$ are the distinct signatures  synchrotron 
emission at the CENBOL. The bump at the lower energy is due to the thermal electrons 
and the bump at the higher energy is due to the power-law electrons.
The photons around $Log(\nu)\sim 18-22$ are produced mainly due to Comptonization of 
the thermal and non-thermal electrons. When it is dominated by the emission from the non-thermal electrons,
it also carries the signature of the CENBOL as well.  Notice in curve 7, how the spectral slope $\alpha$
changes from that of the thermal photons $\alpha_{th}$,
to that of the non-thermal photons $\alpha_{nth}$. The major difference between 
the two cases is that in Fig. 2a, the spectrum is softer since the
electron number density is higher for the low mass black hole, while 
in Fig. 2b, the spectrum is harder as the electron number density 
is lower for the high mass black hole and the electrons are difficult to cool 
by the synchrotron photons emitted by lower density electrons. 

\begin {figure}
\vbox{
\vskip 0.0cm
\hskip 3.0cm
\centerline{
\hskip 0.5cm
\psfig{figure=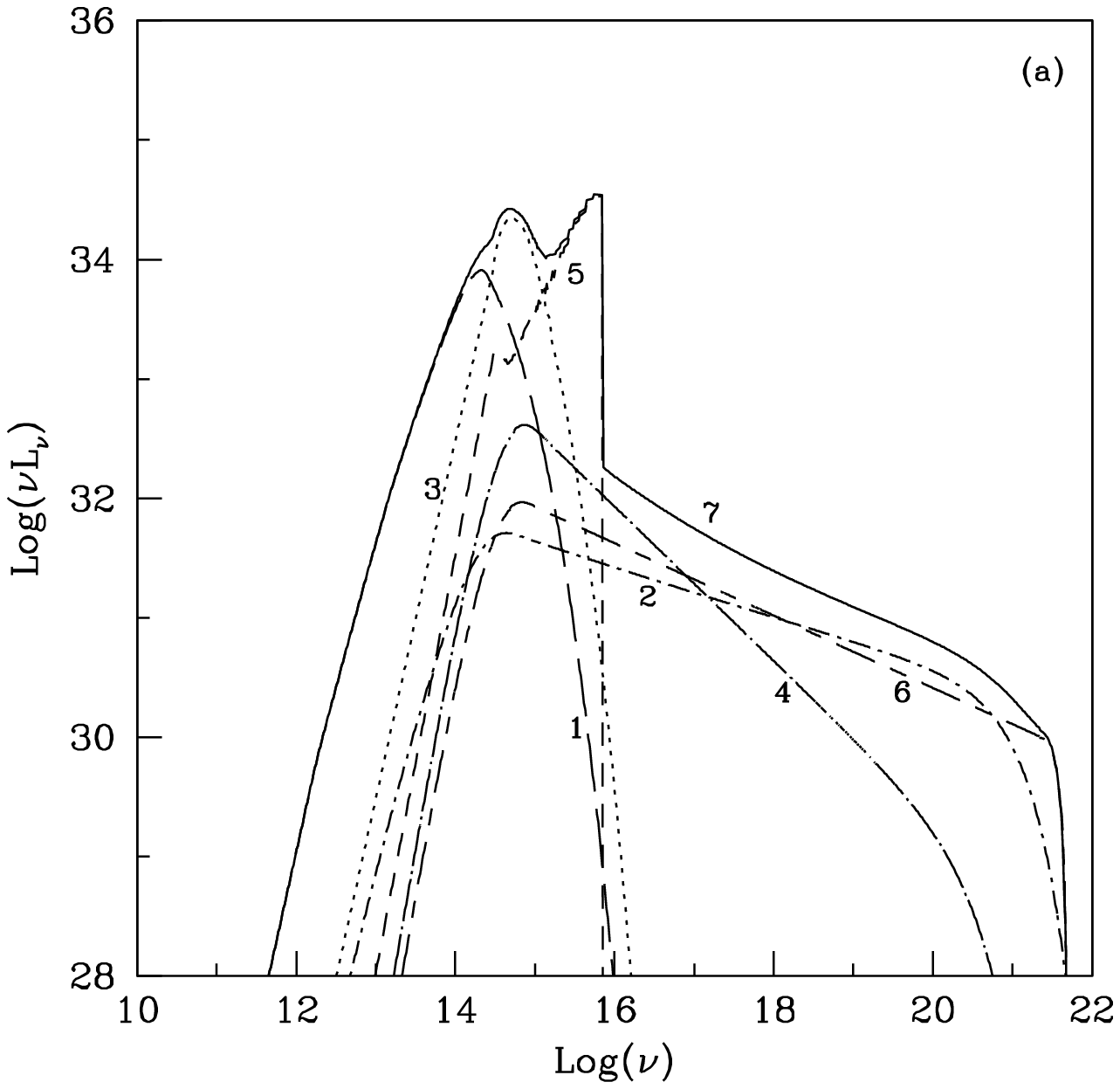,height=8truecm,width=8truecm}
\hskip -1.0cm
\psfig{figure=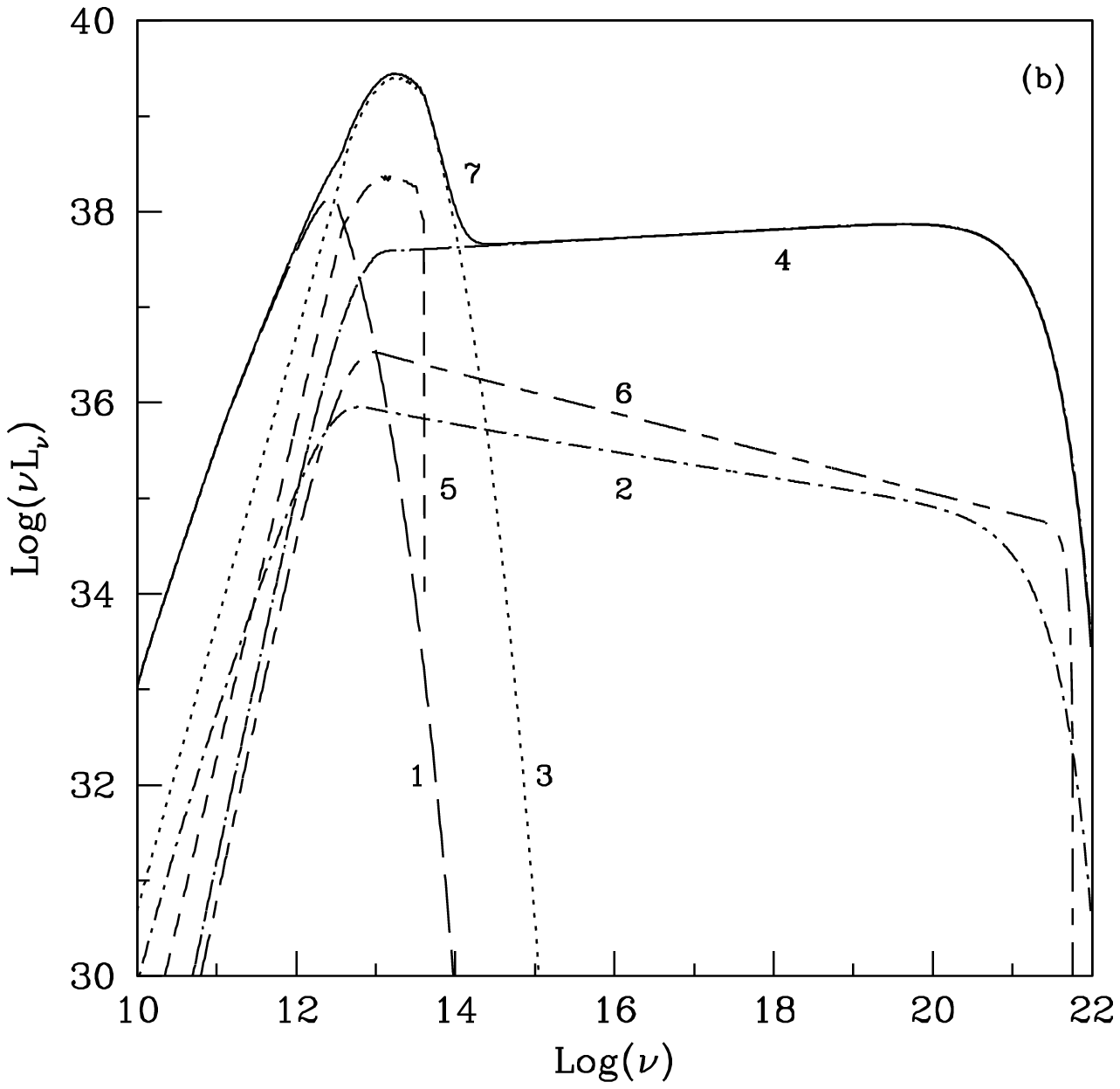,height=8truecm,width=8truecm}}}
\vspace{0.0cm}
\caption{Typical complete spectra from a sub-Keplerian accretion flow with all
the contributions shown for the same parameters as in Fig. 1a except for
the black hole mass: (a) $M=10M_\odot$ and (b) $M=10^6M_\odot$. 
See text for details of the components marked individually. In (a), the spectrum 
is softer and in (b), the spectrum is harder due to lower density. }
\end{figure}

The location of the shock can change depending on the specific angular momentum
and specific energy of the flow. When the shock is located at a larger distance,
much bigger volume of the gas becomes hotter and the spectrum is expected to become
harder. When the shock moves in, it is easy to cool the small sized CENBOL and the
spectrum could become softer. We see this effect in Fig. 3 where we vary the 
shock location. The solid, dotted and the dashed curves denote the cases when 
$x_s=50,\ 25$ and $10$ respectively. Other parameters are  ${\dot m}=0.05$,$\zeta=0.2$.
When the shock location is very close the number of excursion by the electrons
before escaping is fewer and hence $\gamma_{max}$ is lower (see, eq. 37). This
causes the higher energy bump to disappear. 
When the shock is farther away, the CENBOL radiation dominates the spectrum.
Effectively, these three cases can be thought of having an 
extremely rotating Kerr black hole with contra-rotating disk, 
a Schwarzschild black hole, and an extremely rotating Kerr black hole with co-rotating 
disk respectively. 

\begin {figure}
\vbox{
\vskip -3.0cm
\hskip 0.0cm
\centerline{
\psfig{figure=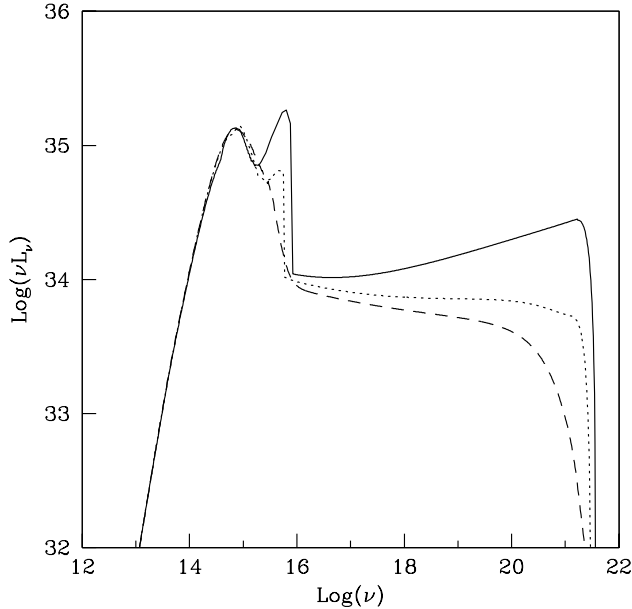,height=13truecm,width=13truecm}}}
\vspace{0.0cm}
\caption{Variation of the total emitted spectrum when the shock location $x_s$ is varied. The solid, 
dotted and dashed curves are drawn for $x_s=50,\ 25,\ 10$ respectively. The spectrum becomes softer 
when $x_s$ is reduced.}
\end{figure}

We show now that the extent to which the pre-shock and the post-shock flows dominate the spectral 
feature depends on the compression ratio of the shock. It is well known (Landau \& Lifshitz, 1956) that 
when the shock is strong, its compression ratio $R$ is given by,
$$
R_{strong}=\frac{\gamma_{pol}+1}{\gamma_{pol}-1},
\eqno{(38)}
$$
where, $\gamma_{pol}$ is the polytropic index of the equation of state $P \propto \rho^{\gamma_{pol}}$.
In the case of the mono-atomic gas, $\gamma_{pol}= 5/3$ and $R_{strong}=4$. However,
for weak shocks, the strength may be $R\sim 2-3$. For the weakest shock (i.e.,
no shock), $R=1$. For an extremely relativistic flow, $\gamma_{pol}=4/3$ and the shock may have 
a strength up to $7$. In Fig. 4, we show the variation of the spectrum 
as a function of the compression ratio $R$. The accretion rate is ${\dot m}=0.5$ and 
$\zeta=0.2$. Solid curves are drawn for  $R=4$, the dotted curves are for $R=2.6$ and 
the dashed curves are for $R=1.5$. For a weak shock, the jump in density $\rho_+ = R\rho_-$, 
where $+$ and $-$ denote the post and pre-shock values respectively, at the 
shock is not so high and post-shock flow does not leave any distinct signature
(long-dashed). Exactly opposite phenomenon occurs for a strong shock (solid). The accretion 
rate being high, it is difficult to cool the electrons, especially when the compression 
ratio is large, i.e., the electrons in the CENBOL are hotter. As a result, the spectrum  is softer
for $R=1.5$ and harder for $R=4$. 

\begin{figure}
\vbox{
\vskip -3.0cm
\centerline{
\hskip 2.5cm
\psfig{figure=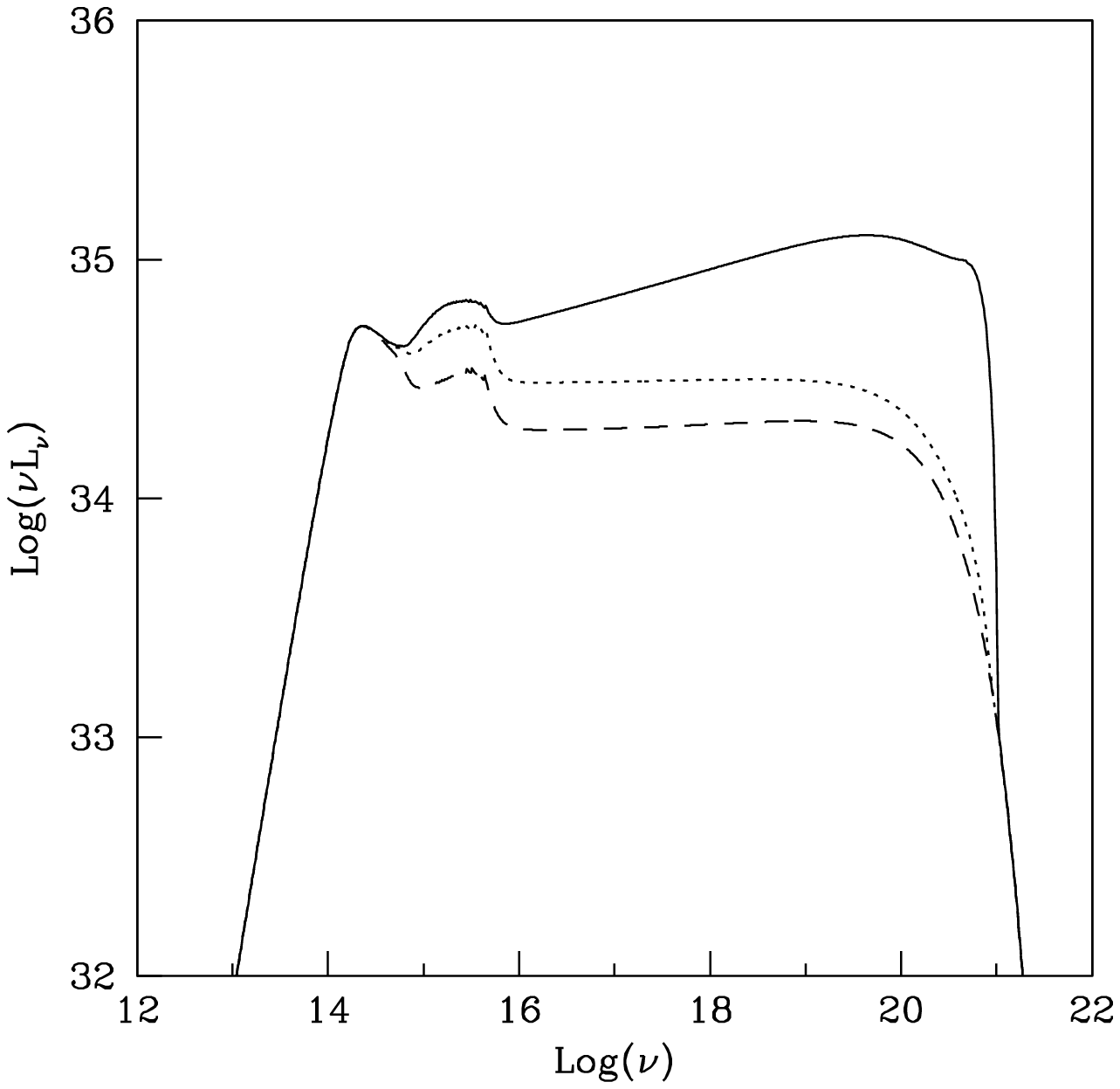,height=11truecm,width=11truecm}}}
\vspace{0.0cm}
\caption{Variation of the spectrum when the compression ratio $R$ is varied.
Solid, dotted and dashed curves are for $R=4.0,\ 2.6\ and\ 1.5$ respectively.
In general, the spectrum hardens with the increase of the compression ratio.}
\end{figure}

In Fig. 5, we show the variation of the spectrum when ${\dot m}$ is varied. 
Here, $\zeta=0.2$ is used. A higher accretion rate is expected to increase 
the density of the flow which enhances the magnetic field and therefore the
synchrotron cooling is also increased. However, the cooling is not sufficient 
when the optical depth enhancement is also taken into account. Thus, the spectrum 
becomes harder despite of this cooling. The Comptonized spectrum becomes harder. 
At lower accretion rate, it is easy to cool the flow and the spectrum becomes softer. 
The curves are drawn for: ${\dot m}=1.0$ (solid), $0.5$ (dotted) and $0.1$ (dashed). 

As we discussed earlier, not all the accreting matter actually passes through the 
shock. Low angular momentum matter (for a given specific energy) produces lesser jet and 
if shock exists, most of the matter passes through the shock located closer to the hole. 
High angular momentum inflow produces profuse outflows because the CENBOL size becomes very high. 
The fraction of electrons which have power-law distribution $\zeta$ thus varies from 
$0$ to $1$ depending on the flow parameters. In Fig. 6, we show the variation of 
the outgoing spectrum as a function of $\zeta$. The dashed curve is for $\zeta=0$ 
and the dotted and solid curves are for $\zeta=0.5$ and $\zeta=0.7$ respectively. 
We use ${\dot m}=0.05$ and $x_s=25$. The spectrum changes from soft to hard with the 
increase of $\zeta$. Note that the bump around $\nu\sim 10^{18}$Hz is also 
weaker for smaller $\zeta$.

\begin {figure}
\vbox{
\vskip -3.0cm
\centerline{
\hskip 0.0cm
\psfig{figure=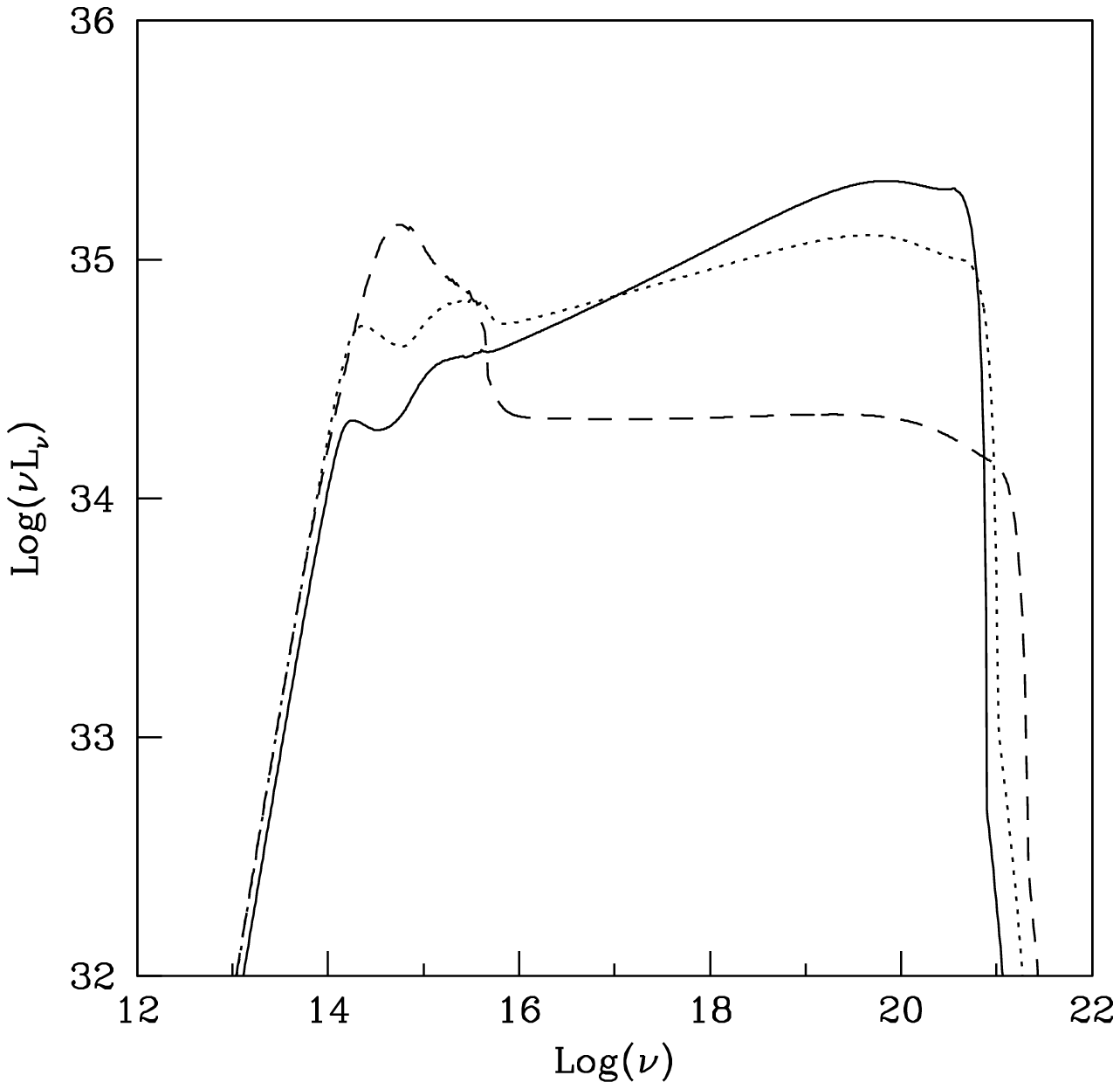,height=13truecm,width=13truecm}}}
\vspace{0.0cm}
\caption{Variation of the spectra with the dimensionless accretion rate. Solid, 
dotted and dashed curved are for ${\dot m}=1.0,\ 0.5$ and $0.1$ respectively. 
The spectrum becomes harder when the accretion rate is increased due to poor cooling
effects.}
\end{figure}

\begin {figure}
\vbox{
\vskip -3.0cm
\centerline{
\hskip 2.5cm
\psfig{figure=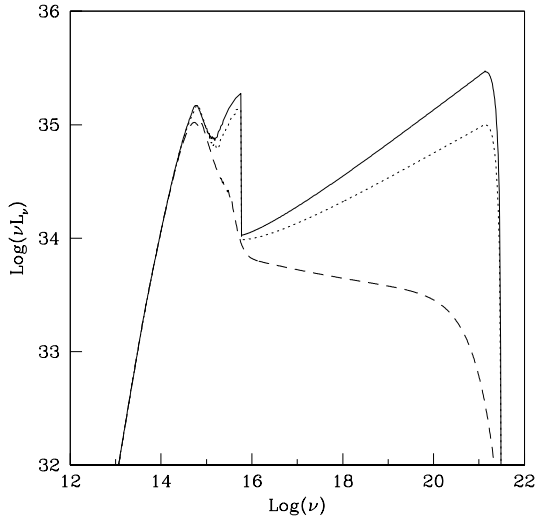,height=11truecm,width=11truecm}}}
\vspace{0.0cm}
\caption{Variation of the spectrum as a function of the percentage of 
electrons ($\zeta$) which passes through an accretion shock. The 
dashed, dotted and solid curves are drawn for $\zeta=0.0,\ 0.5\ and\ 0.7$ 
respectively. The spectrum hardens with the increase in $\zeta$.}
\end{figure}

\begin {figure}
\vbox{
\vskip -3.0cm
\hskip 0.0cm
\centerline{
\psfig{figure=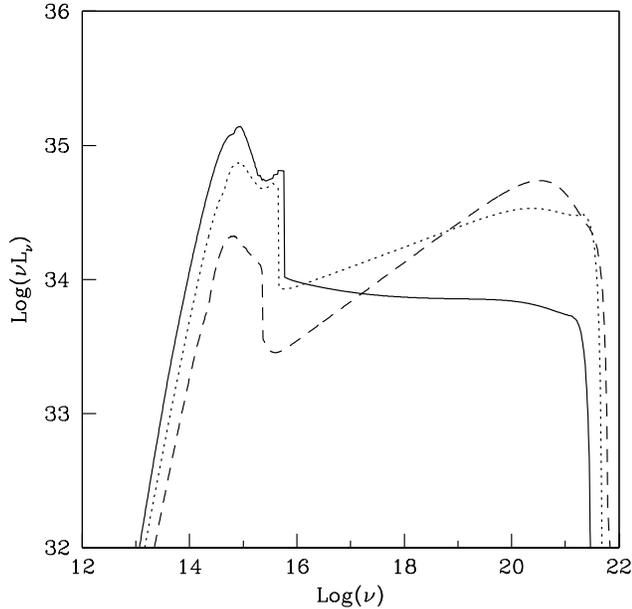,height=13truecm,width=13truecm}}}
\vspace{0.0cm}
\caption{
Variation of the emitted spectrum when the ratio $\beta$ between the 
magnetic energy density and the gravitational energy density is varied. 
Solid, dotted and dashed curves are for $\beta=1,\ 0.1\ and\ 0.01$
respectively. The spectrum softens when the magnetic energy is increased.}
\end{figure}

\begin {figure}
\vbox{
\vskip -2.0cm
\centerline{
\hskip 5.0cm
\psfig{figure=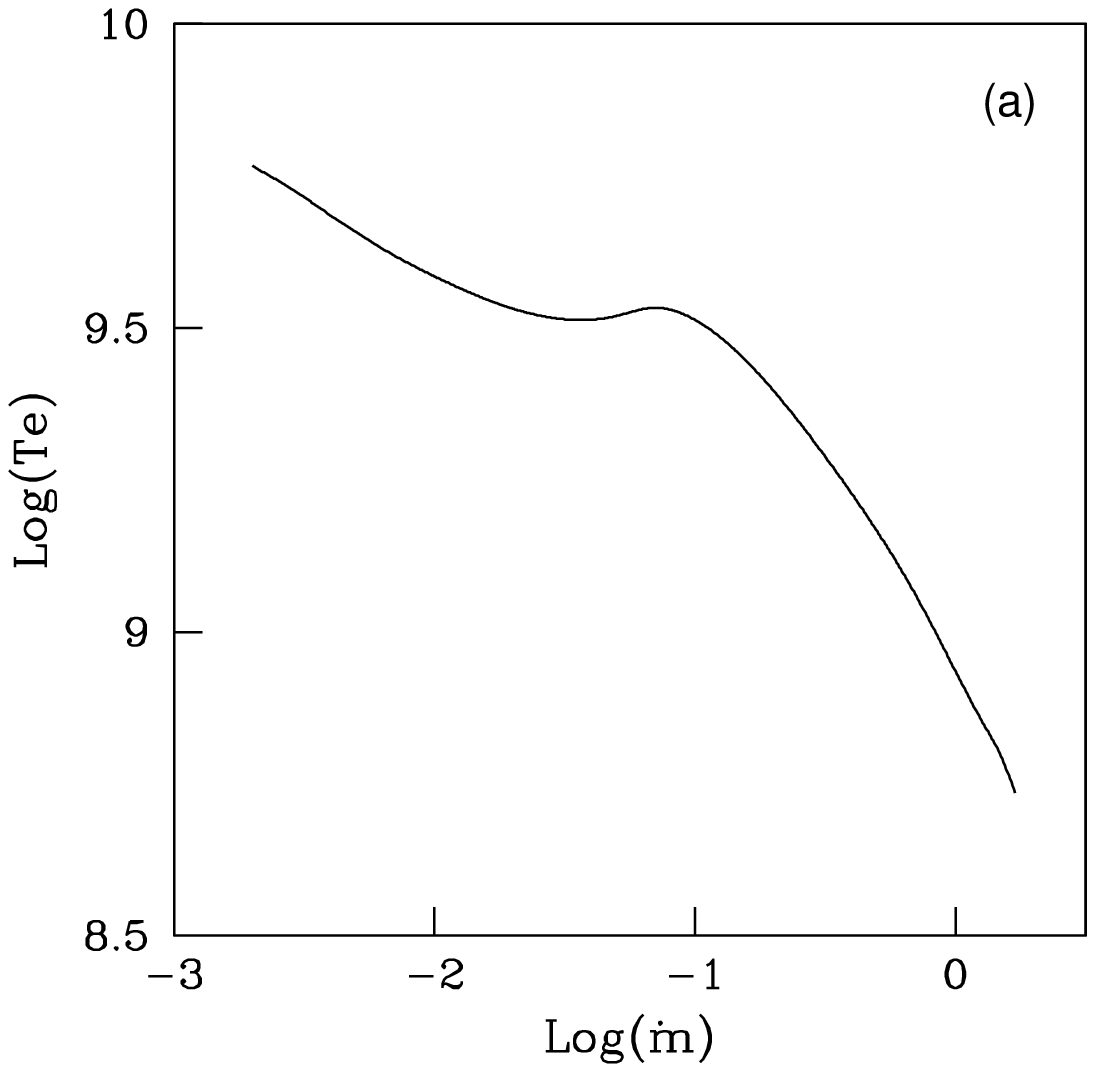,height=12truecm,width=12truecm}}
\vskip -4.0cm
\centerline{
\hskip 5.0cm
\psfig{figure=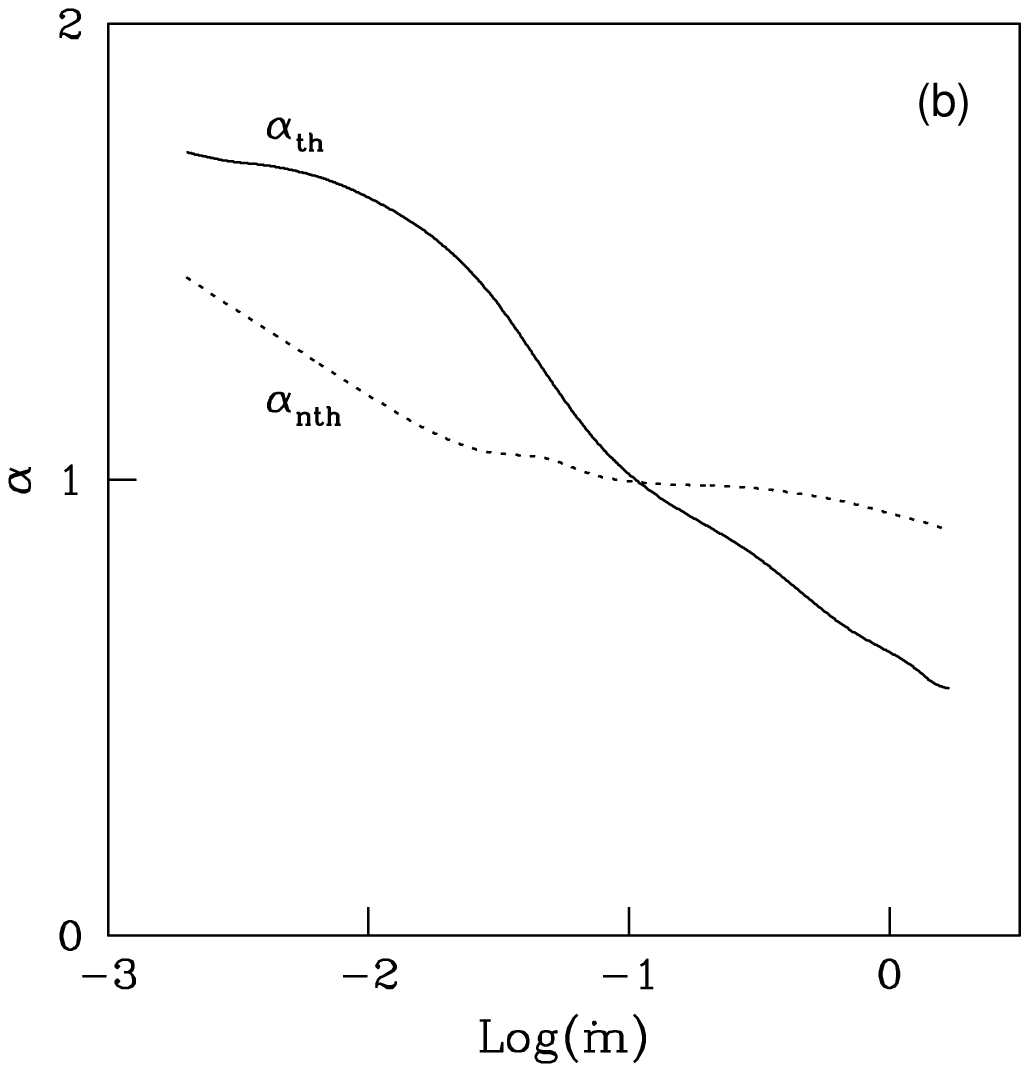,height=12truecm,width=12truecm}}}
\vspace{0.0cm}
\caption{Variation of the (a) average electron temperature and (b) thermal (solid) and non-thermal 
(dotted) spectral indices as functions of the mass accretion rate.}
\end{figure}

In all the cases mentioned above, we have assumed that the stochastic magnetic field 
is amplified sufficiently to become equipartition value at each radial distance. There are
several reasons in both for and against this argument. In a conical wedge shaped flow as we
used here, $B\sim 1/r^2$ and the magnetic pressure $\sim 1/r^4$ increases much faster compared with 
the gas pressure $P_{gas} \sim 1/r^{5/2}$. Thus, it is not unlikely that 
magnetic field, equipartition with gas pressure at a given outer radius, could 
remain in equipartition in all the radii by appropriately releasing 
magnetic energy. However, the time scale of amplification could be comparable to the 
buoyancy time-scale (Nandi et al. 2001) and it is not necessary that the {\it low angular flow}
would get enough time to be amplified to the equipartition value. In the absence of 
any time-dependent simulation results in this direction, we put the ratio of the
magnetic energy to the gravitational energy, namely, $\beta$ to be a parameter. This enables us to 
understand the importance of the magnetic field. Since we are not considering any Keplerian disk,
the only source of soft photon is the synchrotron radiation, the intensity of which directly
depends on the magnetic field. Figure 7 shows the spectral variation. The solid, dotted and the dashed
curves are drawn for $\beta=1,\ 0.1,$ and $0.01$ respectively. As expected, at low
$\beta$, the spectrum is hard (dashed) and at high $\beta$, the spectrum is soft (solid). Thus,
the spectral transition may also be triggered by the degree of magnetization of the flow. 

While discussing Fig. 5, we have already mentioned that, when the accretion rate is increased 
the cooling takes place but the spectrum is hardened. This behaviour is generally 
seen in Figs. 8(a-b) where we present the variation of (a) the average electron 
temperature and (b) the spectral indices of the thermal ($\alpha_{th}$, solid) and 
($\alpha_{nth}$, dotted) with the mass accretion rate. The parameters are $x_s=35$ and $\zeta=0.1$.
In this case, a bump is seen around ${\dot m} \sim 0.1$ where the electron temperature rises momentarily.
Below this bump, the synchrotron cooling dominates and above this bump the Compton cooling
dominates. Close to the bump these are of comparable magnitude.  
Both the thermal and non-thermal electrons show similar behaviour, namely, hardening with the
increase of the accretion rate. It is to be noted that this is markedly different from CT95 result
where the flow was becoming softer with the accretion rate. This is because, in this paper, we are
plotting the variation with the halo rate (sub-Keplerian flow), while in CT95, it was the Keplerian 
rate that was considered. There too it was found that the spectral index goes down with the increase
in the halo rate. Thus the behaviour in presence of synchrotron radiation (not present in CT95)
is qualitatively similar. 

\section{Discussion and Concluding Remarks}

In this paper, we have highlighted one major issue: to take care of the effect of accretion shocks
on the spectral properties. To our knowledge, no such studies have been done in the literature.
An accretion shock separates a flow into two parts: the pre-shock flow 
is of lower density but of higher volume, while the post-shock flow (CENBOL) is of higher density but of 
lower volume. More importantly, the shock accelerates particles and generate non-thermal electrons.
We consider this generation self-consistently. It is true that we used the shock location
and the compression ratio as independent parameters, but since we did not
explicitly use the angular momentum of the matter, this re-parametrization is totally consistent.
In this way, we have thus incorporated the emission from the Maxwell-Boltzmann electrons and the power-law electrons
as well as Comptonization of these emissions. Since the power-law electrons are present in the CENBOL, 
the radiation from these electrons is also the signature of the CENBOL.

In this paper, we have shown that, generally speaking, the spectral transition from a hard-state to a soft-state 
can be triggered by (a) reducing shock location (Fig. 3), (b) by reducing the compression ratio (Fig. 4),
(c) by reducing the accretion rate (Fig. 5), (d) by reducing power-law electron contents (Fig. 6),
and (e) by increasing the magnetic field strength inside the flow (Fig. 7). Note that, so far, we did not
introduce the soft-photons from a Keplerian disk as was done in CT95. There, we did see that 
for a given rate of the Keplerian flow, if the accretion rate of the halo (sub-Keplerian flow) is increased,
the spectrum becomes harder -- an effect reflected in Fig. 5 above even when synchrotron photons are
present.  In future, we shall include the Keplerian disks as well. We shall also compare our
results with observed spectra in order to pinpoint signatures of shocks in an accretion flow. In 
presence of a Keplerian disk, the soft X-ray bump will be produced at around a few KeV, while for shocks
the bump due to Maxwell-Boltzmann electrons is located in UV while that of the power-law electrons
is located at a few hundred keV to a few MeV. Thus, it is possible that these 
two distinct bumps can be distinguished through observation. 

The outflows produced from a CENBOL itself can have shocks. The shocks
in outflows are located very close to the black holes. Thus, a pre-shock flow has a very little
volume. Post-shock outflows cool off rapidly and the cooler electrons  ($T<10^6$K) can also
produce a low-energy, power-law spectrum. On the other hand, there could be several shocks
in the jets. Their effects from our CENBOL would be discussed elsewhere.

The authors thank the anonymous referee for suggestions leading to improvements of the paper. This work is 
partly supported by a grant from  the RESPOND project of the Indian Space Research Organization.

\end{document}